\documentclass[12pt]{article}
\usepackage[utf8]{inputenc}
\usepackage{amsmath,amssymb,graphicx,epsfig,color,float,cancel,cite,textcomp}
\usepackage[titletoc,title]{appendix} 
\usepackage{longtable}
\usepackage{caption}
\usepackage[export]{adjustbox}
\usepackage{multicol,tabu,multirow}
\usepackage{arydshln}
\usepackage{hyperref}
\usepackage{slashed}                                                                                                                                                                                                                                                                                                                                                                                                                                                                                                                                                                                                                                                                                                                                                                                                                                                                                                                                                                                                                                                                                                
\numberwithin{equation}{section}

\def\nn{\nonumber}

\def\nn{\nonumber}
\def\bea{\begin{eqnarray}}
\def\eea{\end{eqnarray}}

\newcommand{\beq}{\begin{equation}}
\newcommand{\eeq}{\end{equation}}
\newcommand{\bseq}{\begin{subequations}}
\newcommand{\eseq}{\end{subequations}}
\newcommand{\beqa}{\begin{eqnarray}}
\newcommand{\eeqa}{\end{eqnarray}}
\newcommand{\half}{\frac{1}{2}}

\newcommand{\del}{\partial}

\newcommand{\tr}{\textrm{tr}}

\def\ThZ{$1$}
\def\ThO{$\omega$}
\def\ThTw{$\omega^2$}

\def\FZ{$\hphantom{-}1$}
\def\FO{$\hphantom{-}i$}
\def\FTw{$-1$}
\def\FTh{$-i$}
\begin{document}
\thispagestyle{empty}
\begin{center}
\vspace*{1cm}
{\LARGE\bf Note on  gauge and gravitational anomalies of discrete $Z_N$ symmetries \\}
\bigskip
{\large Pritibhajan Byakti}\,$^{a,1}$, \, \,
{\large Diptimoy Ghosh}\,$^{b,2}$,\, \, 
{\large Tarun Sharma}\,$^{c,3}$
\\
\bigskip 
{\small
$^a$ Department of Theoretical Physics, Indian Association for the Cultivation of Science,
\\ 2A \& 2B, Raja S.C. Mullick Road, Jadavpur, Kolkata 700 032, India. \\[2mm]
$^b$ Department of Particle Physics and Astrophysics, Weizmann Institute of Science, 
\\ Rehovot 76100, Israel. \\[2mm]
$^c$ Mandelstam  Institute  for  Theoretical  Physics,  School  of  Physics,  and  National  Institute
for  Theoretical  Physics,  University  of  the  Witwatersrand,  \\ 
Johannesburg,  WITS  2050, South Africa.
}
\end{center}
\bigskip 
\vspace*{0.4cm}
\begin{center} 
{\Large\bf Abstract} 
\end{center}
\vspace*{-0.35in}
\begin{quotation}
\noindent  In this note, we discuss the consistency conditions which a discrete $Z_N$ symmetry should satisfy 
in order that it is not violated by gauge and gravitational instantons.  As examples, we enlist all the $Z_N$ 
${\cal R}$-symmetries as well as non-${\cal R}$ $Z_N$ symmetries (N=2,3,4) in the minimally 
supersymmetric standard model (MSSM) that are free from gauge and gravitational anomalies. 
We show that there exists non-anomalous discrete symmetries that forbid Baryon number violation up to 
dimension 6 level (in superspace). We also observe that there exists no non-anomalous $Z_3$ ${\cal R}$-symmetry 
in the MSSM. Furthermore, we point out that in a theory with one Majorana spin 3/2 gravitino, a large class of $Z_4$ 
${\cal R}$-symmetries are violated in the presence of Eguchi-Hanson (EH) gravitational instanton. This is also in general true 
for higher $Z_N$ ${\cal R}$-symmetries. We also notice that in 4 dimensional ${\cal N}=1$ supergravity, the global $U(1)$ ${\cal R}$-symmetry 
is always violated by the EH instanton irrespective of the matter content of the theory.
\end{quotation}
\bigskip
%
%
\vfill
%
%
\bigskip
\hrule
\vspace*{-0.1in}
\hspace*{0.1cm}$^1$ tppb@iacs.res.in ~~~
$^2$ diptimoy.ghosh@weizmann.ac.il ~~~
$^3$ tarun.sharma@wits.ac.za

\newpage 
\tableofcontents
\section{Introduction}

Symmetry has been a central concept in the formulation of the fundamental laws of physics. While the concept of 
local gauge invariance has played a pivotal role in elementary particle physics, providing for example, the basis for 
the extremely successful Standard Model (SM), global symmetries have also been used extensively in building models 
of particle physics.
However, there exists a widespread notion that all global symmetries, both continuous and discrete, are violated 
by gravitational effects \cite{Krauss:1988zc,Kallosh:1995hi,Banks:2006mm,Banks:2010zn}.
Thus, it is perhaps best to not impose any global symmetry on the Lagrangian, and a model  should only be 
constructed based on gauge symmetries\footnote{Global symmetries can certainly emerge as accidental symmetries 
at long distance.}
\footnote{Also note that, global discrete symmetries, if spontaneously 
broken, lead to stable domain walls which may pose serious difficulties in cosmology \cite{Preskill:1991kd}.}.

However, gauging continuous global symmetries has  dynamical consequences, the most important being the 
existence of gauge fields. This often creates problems for phenomenology. On the other hand, gauging discrete 
symmetries does not introduce any such problem, and this is why imposing discrete symmetries has proved to be 
a very useful tool in building viable models of particle physics.

At first sight, the very idea of gauging a discrete symmetry looks rather unclear. This is because, the difference 
between a gauged discrete symmetry and its global version looks artificial in the sense that, at least to our knowledge, there 
is no clear observable consequence of gauging a global discrete symmetry.\footnote{The state space for a global 
$Z_N$ symmetry is definitely different form that of a gauged $Z_N$ symmetry. So, in principle, one might be able to 
find observable that distinguishes a global $Z_N$ symmetry to a gauged one.} Therefore, 
when we talk about gauging a discrete symmetry, we only mean that the discrete symmetry should be part of a 
gauge symmetry in the Ultra Violet (UV).  Being part of a gauge symmetry, it is immediately obvious that the 
discrete symmetry\footnote{In this work, we will only worry about discrete $Z_N$ symmetries.} should not have 
quantum anomalies with itself, other gauge symmetries or gravity. 
In this note, we investigate this question in a systematic way, with a particular emphasis on the mixed anomaly 
with gravity. 
We apply these anomaly constraints to the Minimally Supersymmetric Standard Model (MSSM), and show that a 
large class of discrete $Z_N$ symmetries in particular, $Z_N$ ${\mathcal R}$ symmetries are ruled out as gauge 
symmetries if breaking by gravitational instantons (in particular, the Eguchi-Hanson instanton) are taken into 
account.

The criteria for discrete symmetries to be non-anomalous were first studied in \cite{Ibanez:1991hv} (for later studies, 
see \cite{Ibanez:1991pr,Ibanez:1992ji,Martin:1992mq,Martin:1993zr,Kurosawa:2001iq,Dreiner:2005rd,Araki:2008ek,Lee:2010gv,
BerasaluceGonzalez:2011wy,Lee:2011dya,Dreiner:2012ae,Honecker:2013sww,Chen:2013dpa,Chen:2014gua,
Mayrhofer:2014laa,Chen:2015aba}). Their analysis involved 
embedding of the discrete symmetry in a continuous group which was assumed to be spontaneously broken at some high 
energy scale. The continuous symmetry group was then required to be non-anomalous (including their cubic anomalies). 
However, as pointed out in \cite{Banks:1991xj}, part of the constraints obtained in \cite{Ibanez:1991hv} in particular, the ones 
non-linear in the discrete charges, relied on the way the discrete symmetry was embedded in a particular high energy theory. 
As a consequence, these constraints are dependent on the UV physics and are not necessarily required to 
be satisfied in the low energy theory. We do not use these UV dependent constraints in this note. Rather, we only use 
constraints that can be derived from the low energy theory. To be precise, we first embed the discrete symmetry in a 
U(1) group (we call this U(1) to be the mother group) and check whether it has mixed anomalies with the SM gauge 
symmetries and gravity. If the mother U(1) turns out to have certain mixed anomalies then we look at the instanton processes 
that violates the U(1) symmetry and make sure that those instantons do not violate the discrete subgroup of our interest. 

The paper is organised as follows. In the next section, as a warm-up, we consider the case of U(1) ${\mathcal R}$ symmetry 
in the MSSM, and investigate whether there exists one which has no mixed anomaly with the gauge symmetry. 
We then pose the question as to whether a discrete subgroup of U(1) ${\mathcal R}$ symmetry can be non-anomalous. 
In section~\ref{gauge-anomaly},  we show how anomalies in discrete $Z_N$ symmetries can be calculated. We discuss 
anomalous violation of discrete symmetries both by the gauge and gravitational instantons. We show examples of 
non-anomalous $Z_2$, $Z_3$ and $Z_4$ symmetries in the renormalisable MSSM as well as MSSM with higher dimensional 
operators  in section~\ref{examples}. We summarise our findings in section~\ref{summary}.

\section{U(1) ${\mathcal R}$ symmetry in the MSSM}

The MSSM superpotential can be written as a sum of three pieces,
\bea
\label{mssmsp}
W_{\rm MSSM} = W_0 + W_{\slashed{L}} + W_{\slashed{B}} 
\eea

where, $W_0$, $W_{\slashed{L}}$ and  $W_{\slashed{B}}$ are given by, 
\bea 
W_0 &=& Y^u_{ij} Q_i U^c_j H_u + Y^d_{ij} Q_i D^c_j H_d + \mu H_u H_d + Y^\ell _{ij} L_i E^c_j H_d \\
W_{\slashed{L}} &=& \dfrac{1}{2} \lambda_{ijk} L_i L_j E^c_k + \lambda^{'}_{ijk} L_i Q_j D^c_k + \tilde{\mu}_i L_i H_u \\
W_{\slashed{B}} &=& \dfrac{1}{2} \lambda^{''}_{ijk} U^c_i D^c_j D^c_k
\eea

For convenience, the gauge, $L$ and $B$ quantum numbers of all the MSSM superfields are shown in table \ref{mssmf}.
Clearly, $W_0$ does not violate Lepton number ({$L$}) or Baryon number ($B$), however,  $W_{\slashed{L}}$ and 
$W_{\slashed{B}}$ violate $L$ and $B$ respectively.

\begin{table}[h]
\centering
\begin{tabular}{|c|c|c|c|c|c|}
\hline 
    & SU(3)   & SU(2)  & Y/2 & B  & L \\
\hline 
$Q$   & 3       & 2      & 1/6  & 1/3  & 0 \\
\hline
$U^c$   & $\bar{3}$ & 1      & -2/3 & -1/3 & 0 \\
\hline
$D^c$   & $\bar{3}$ & 1      & 1/3  & -1/3 & 0 \\
\hline
$L$   & 1       & 2      & -1/2 & 0  & 1 \\
\hline
$E^c$   & 1       & 1      & 1    & 0  & -1 \\
\hline
$H_u$ & 1       & 2      & 1/2  & 0  & 0 \\
\hline
$H_d$ & 1       & 2      & -1/2  & 0  & 0 \\
\hline
\end{tabular}
\caption{MSSM matter superfields and their gauge, $B$ and $L$ charges \label{mssmf}}
\end{table}

The superpotential of Eq.\eqref{mssmsp} has many choices of ${\mathcal R}$ symmetries at the 
classical level. For example, let us consider the $W_0$ piece for simplicity. In order to get the allowed ${\mathcal R}$-charges, 
one has to solve the following simultaneous equations
\bea
\label{Reqn1}
R_Q +R_{U^c} + R_{H_u}   & = 2 R_\theta \\
R_Q+ R_{D^c} + R_{H_d}   & = 2  R_\theta \\
R_{H_u} + R_{H_d}        &  = 2  R_\theta \\
R_L +R_{E^c} + R_{H_d}   &  = 2  R_\theta
\eea
which have four parameter worth of solutions, given by
\bea
R_{D^c} &=& 2  R_\theta - 2 R_Q - R_{U^c} \label{solReqn1-1}\\
R_{E^c} &=& 2  R_\theta - R_L - R_Q - R_{U^c} \\
R_{H_u} &=&  2 R_\theta - R_Q - R_{U^c} \\
R_{H_d} &=& R_Q + R_{U^c} \label{solReqn1-4}
\eea
One can now check whether there exists any choice of ${\mathcal R}$-symmetry that is free from mixed anomalies 
with the gauge symmetries. 
The anomaly coefficients for the mixed \mbox{${\mathcal R}$-$SU(3)$-$SU(3)$}, \mbox{${\mathcal R}$-$SU(2)$-$SU(2)$} 
and \mbox{${\mathcal R}$-$U(1)_{Y/2}$-$U(1)_{Y/2}$} anomalies are given by, 
\bea
{\mathcal A}_3 & = {\rm Tr}\left(\{T^a_{\small SU(3)},T^b_{\small SU(3)}\}  T_{\mathcal R}\right) = & 
3 \left( 2 q_Q + q_{U^c} + q_{D^c} \right) + 6 R_\theta \label{su3-anomaly} \\
{\mathcal A}_2 & = {\rm Tr}\left(\{T^a_{\small SU(2)},T^b_{\small SU(2)}\}  T_{\mathcal R}\right) = &
3 \left( 3 q_Q + q_L \right) + q_{H_u} + q_{H_d} + 4 R_\theta \label{su2-anomaly} \\
{\mathcal A}_1 & = ~~ {\rm Tr}\left(\{T^a_{\small Y/2},T^b_{\small Y/2}\}  T_{\mathcal R}\right) ~~ = &
  3 \left( 6 q_Q (1/6)^2 + 3 q_{U^c}(2/3)^2 + 3 q_{D^c} (1/3)^2 + 2 q_L (1/2)^2 + q_{E^c} \right) \nonumber \\ 
&  & + 2q_{H_u}(1/2)^2 +   2q_{H_d} (-1/2)^2 \label{Y-anomaly}
\eea
Here, $q_i$ are the ${\mathcal R}$ charges of the Weyl fermions inside the superfields 
i.e., $q_i = R_i -  R_\theta$. The constant 
factors $6 R_\theta$ and $4 R_\theta$ are due to the gauginos.  
We have used the normalisation of the SU(N) generators in the fundamental 
representation such that 
${\rm Tr}(T^a T^b) = \dfrac{1}{2} \delta^{ab}$. 
Using the solutions from Eq.\eqref{solReqn1-1}-\eqref{solReqn1-4}, we get 
\bea
{\mathcal A}_3 &=& 0 \\
{\mathcal A}_2 &=& -8 R_\theta + 3 R_L + 9 R_Q \label{A2sol} \\
{\mathcal A}_1 &=& -4  R_\theta - 3 R_L - 9 R_Q \label{A1sol}
\eea
It is clear that ${\mathcal A}_2 = 0$ and ${\mathcal A}_1 = 0$ can not be satisfied 
simultaneously (because, by definition, $R_\theta \neq 0$).  Thus, we conclude that 
there is no non-anomalous ${\mathcal R}$ symmetry of the superpotential $W_0$. However, there may exist some 
$Z_N$ subgroup of the $U(1)$ ${\mathcal R}$ symmetry that is non-anomalous\footnote{Note that, the U(1) 
${\mathcal R}$ symmetry must anyway be broken in order to generate Majorana gaugino masses.}. We discuss this in the next subsection. 

\section{Anomalies in discrete symmetries}
\label{gauge-anomaly}

Let us consider an $SU(N_g)$ gauge theory with left handed chiral fermions 
in representation $\{{\cal R}_i\}$. We would like to calculate the anomaly in a discrete $Z_N$ 
symmetry which is a subgroup of a  continuous $U(1)$ symmetry defined by the 
transformation $\psi_i \rightarrow e^{i\alpha \, q_i} \psi_i$, where the parameter 
$\alpha$ takes values on $S^1$ of length $2\pi$. The discrete $Z_N$ subgroup is generated 
by $\alpha= \frac{2\pi m}{N}$ with $m \in \{ 0,1,2 \ldots N-1 \}$ and all the fields carry 
integer valued $U(1)$ charges $q_i \in \{0,1 \dots N-1\}$\footnote{The integer 
quantization of $q_i$ follows from the compactness of the group. Further, since we are 
interested only in the discrete $Z_N$ subgroup, $q_i \equiv q_i +N$. So 
we can restrict to $q_i \in \{ 0,1,\ldots N-1 \}$. If one uniformly rescales all the charges by a real number 
$q_0$, then the length of $S^1$, on which the $U(1)$ parameter $\alpha$ takes values, has to be 
rescaled by $q_0^{-1}$ for consistency of the charge quantization condition. This rescaling e.g. can be 
used to set the R-charge of superspace coordinate to the standard value of unity.}. 

In the path integral formulation, the anomaly is 
understood as a consequence of the non-invariance of the fermionic path integral measure.
Assuming that the transformation $\psi_i \rightarrow e^{i\frac{2\pi}{N} q_i} \psi_i$ is a symmetry 
of the classical action, the change in the path integral corresponding to a $Z_N$-$SU(N_g)^2$ 
anomaly can be formally written as, 
\bea
Z[\psi, \psi^\dagger, A_\mu] &=& \int [d\psi] [d\psi^\dagger][dA_\mu] \, \exp \left(i S[\psi, \psi^\dagger, A_\mu] \right) \\
&& \hspace*{-3cm}\rightarrow  \int [d\psi] [d\psi^\dagger][dA_\mu] \, \exp \left(i S[\psi, \psi^\dagger, A_\mu] \right) \, 
\exp \left( i \frac{2\pi}{N}  \sum_{j} q_j  I_{1/2}({\cal R}_j)  \right) \label{tmp1}
\eea
where, the summation runs over all the left handed fermions, and $I_{1/2}({\cal R}_j)$ 
is the Index of the covariant Dirac operator in representation ${\cal R}_j$, and is given by
(see appendix~\ref{Fujikawa} for more details)
\beq\begin{split}
I({\cal R}_j) &= -2T({\cal R}_j) \\
\end{split}\eeq 
in the background of a minimal $SU(N_g)$ gauge instanton (i.e. winding number 1). 
Hence, the discrete $Z_N$ symmetry will be non-anomalous if the quantity $\sum_j q_j ~2T({\cal R}_j)$ 
is an integer multiple of $N$.\footnote{To give some examples, 
in the SM, both the Baryon number symmetry ( $U(1)_B$ ) as well as the Lepton number 
symmetry ( $U(1)_L$ ) have $U(1)-SU(2)-SU(2)$ anomalies. The quantity $\sum_i q_i \, 2T(r_i)$ 
in both the cases turns out to be equal to the number of fermion generations, $n_g$, 
which is 3 in the SM. This means that the $Z_3$ subgroups of  $U(1)_B$ and  $U(1)_L$ 
are not anomalous. In other words, in any process the violation of Baryon or Lepton 
numbers should satisfy $\Delta B, \Delta L = \text{Integer} \times 3$. Hence, such processes 
neither leads to proton decay nor neutrino-less double beta decay (or Majorana neutrino mass). 
Moreover, the combination $U(1)_{B-L}$ is also not anomalous. It would be interesting to check 
whether these $Z_3$ symmetries have gravitational anomalies of not. 
We will comment on that in the next section where we discuss gravitational anomalies.}

In order to compute the mixed anomaly with Hypercharge the relevant anomaly coefficient is 
$\sum_j 2 q_j (Y_j/2)^2$ which is not invariant under ${\rm Exp}[(2\pi/N) q_j] \to {\rm Exp}[(2\pi/N)(q_j + mN)] (m \in \text{Integers})$. 
As for a $Z_N$ symmetry, the discrete charges do not change under the transformation  $q_j \to q_j + mN$, this 
leads to an ambiguity in the calculation of the anomaly. Hence, we do not consider the mixed anomaly with Hypercharge. 
Note that, there is no such ambiguity for the mixed anomaly with $SU(N)$ because $2T({\cal R}_j)$ is an integer 
in that case.

We can now apply this procedure to Eqs.\eqref{solReqn1-1}-\eqref{solReqn1-4}. 
As an example, we can ask whether there exists a $Z_2$ subgroup that is non-anomalous. A non-anomalous $Z_2$ 
symmetry requires ${\mathcal A}_2$ (see Eq.\eqref{A2sol}) to be integer multiple of 2 (${\mathcal A}_3$ is automatically 
zero, and as discussed before, we are not considering the mixed anomaly with Hypercharge). This means that the 
combination $3R_L + 9 R_Q  - 8 R_\theta$ must be integer multiple of 2, which can be easily satisfied 
(by setting either [$R_L = R_Q = 0, R_\theta = 0 \text{ or } 1$] or [$R_L = R_Q = 1, R_\theta = 0 \text{ or } 1$]). Hence, indeed there are 
$Z_2$ subgroups of ${\mathcal R}$-symmetry with the charge assignments of Eq.\eqref{solReqn1-1}-\eqref{solReqn1-4} 
that are non-anomalous.

\subsection{Gravitational anomaly}

In a theory with dynamical gravity, there are additional consistency constraints for 
gauging a global symmetry from the requirement of mixed 
gravitational anomaly cancellation. In analogy with gauge theories, the anomalous 
violation of global symmetries receive contribution from Gravitational instantons
which are finite action solutions of classical euclidean vacuum Einstein 
equations with positive definite metric (with or without the cosmological constant term). 
Depending on the sign of the cosmological constant, a variety of gravitational instanton 
solutions are known (see \cite{Eguchi:1980jx} for a review), a complete classification 
analogous to gauge theories \cite{Atiyah:1978ri} is still lacking though.

As discussed in \cite{Romer:1977qq}, the violation 
of global symmetry by gravitational instanton is also, as in the case of usual gauge 
instantons, due to the asymmetry of zero modes of positive and negative chirality 
of spin 1/2 and spin 3/2 fields, in other words via the corresponding Index. 
The anomalous variation of the path integral of a set of left handed Weyl fermions 
under a discrete $Z_N$ symmetry variation 
in a gravitational instanton background (treated as fixed background) can be written 
in terms of the Indices of the spin 1/2 Dirac operator and spin 3/2 Rarita-Schwinger 
operator  as follows 
\beq\begin{split}
Z[g_{ab}] &= \int  \prod_j [d\psi_j d\psi_j^\dagger] [d\Psi_\mu d\Psi_\mu^\dagger] 
                                                 ~e^{i S[ \psi_j,\Psi_\mu, g_{\mu\nu} ] } \\
 \longrightarrow &  \int \prod_j [d\psi_j d\psi_j^\dagger] [d\Psi_\mu d\Psi_\mu^\dagger] 
                                   ~e^{i S[ \psi_j, \Psi_\mu, g_{\mu\nu} ] }
                                   ~e^{ -i \frac{2\pi}{N} \left[ (\sum_j q_j) I_{1/2} + q_\Psi I_{3/2} \right] } \\
\end{split}\eeq

For Riemannian manifolds without boundaries, according to the Atiyah-Singer Index 
theorem \cite{Atiyah:1963zz},
\bea
I_{3/2} = -21 ~I_{1/2} &=& \frac{-21}{768\pi^2} \int_M \epsilon^{\mu\nu\rho\sigma} 
         R_{\mu\nu\alpha\beta} R^{\alpha\beta}_{~~\rho\sigma}
\eea

In the presence of boundaries, the expressions for $I_{1/2}$ and $I_{3/2}$ also involve 
additional boundary terms 
\cite{Atiyah:1975jf,Atiyah:1976jg,Atiyah:1980jh,Eguchi:1978gw,Nielsen:1978ex,Perry:1978jj,Critchley:1978ki,
Christensen:1978gi,Hawking:1979zs},  
\begin{align}
I_{s} &= -\frac{a_s}{24} \left( P_1[M] - Q_1[\del M] \right) - \half \left( \eta_s + h_s \right), \quad s=1/2, 3/2 \\ 
\textrm{where} & \quad P_1[M] = -\frac{1}{8\pi^2} \int_M \tr(R\wedge R) 
                 = \frac{1}{32\pi^2} \int_M d^4x ~\epsilon^{\mu\nu\rho\sigma} 
                      R_{\mu\nu\alpha\beta} R_{\sigma\rho}^{~~\alpha\beta} ,  \nn \\
& Q_1[\del M] = - \frac{1}{8\pi^2} \int_{\del M} \tr(\theta \wedge R) 
                = \frac{1}{32\pi^2} \int_{\del M} d^3x ~\epsilon^{ijk} \theta_{imn} R_{jk}^{~~mn}, \nn \\ 
& \theta = \omega - \omega_0 \quad \textrm{where, $\omega$ and $\omega_0$ are the spin 
      connection 1-forms for the actual metric} \nn \\ 
&  \hspace{25mm} \textrm{ and corresponding product form metric on $M$.} \nn \\
& \eta_s \rightarrow \textrm{APS $\eta$-invariant}, \nn \\
&  h_s \rightarrow \textrm{dimension of harmonic space of the tangential part of the Dirac (Rarita-Schwinger)} \nn \\ 
&  \text{operator $D_{s=1/2} (D_{s=3/2})$ on $\del M$,} \nn \\
& a_s = (1,-21) \textrm{ for } s=(1/2, 3/2) \textrm{ respectively}. \nn 
\end{align}

In four dimensional theories, the existence of this anomaly can also be understood to 
arise from the triangle diagram with one external gauge boson and two external gravitons 
\cite{Delbourgo:1972xb,Eguchi:1976db,AlvarezGaume:1983ig}\footnote{In order to couple the SM to gravity, the gauge symmetries must be free form 
mixed anomalies with gravity. This imposes the requirement that $\sum Y$ must be zero. Interestingly, this condition is already 
satisfied in the SM because of mixed gauge anomaly cancellation. }. The triangle diagram leads to the anomalous non-conservation of the 
current $J^\mu$ \cite{Nielsen:1978ex,VanNieuwenhuizen:1981ae},
\beq\begin{split}\label{Ganomeq}
D^\mu J_\mu &=\frac{1}{768\pi^2} \left( \sum_{\text{spin 1/2}} q_i - 21 \sum_{\text{spin 3/2}} q_j \right) 
                           \epsilon^{\mu\nu\rho\sigma} R_{\mu\nu\alpha\beta} 
                            R^{\alpha\beta}_{~~\rho\sigma} \\ 
\end{split}\eeq

Various gravitational instanton solutions are known in the literature \cite{Gibbons:1979xm,Eguchi:1980jx}.  
For $\Lambda > 0$, there are only four known examples, all of them being compact 
\cite{Gibbons:1979xm,Eguchi:1976db,Gibbons:1978zy,Page:1979zv}. 
Two of them do not allow spin structures \cite{Hawking:1977ab,Page:1979zv}, and the rest have vanishing index. 
Therefore, these instanton solutions are irrelevant for us. 

In the case of $\Lambda =0$, the only known compact instanton is called $K3$ \cite{hitchin1974}.    
There are also gravitational instantons which possess a self-dual curvature and approach a fiat metric
at infinity. We will discuss them below. 

\subsection{The K3 Instanton}

As stated above, the only solution of Einstein equation with the cosmological constant $\Lambda=0$ which has a compact, 
self dual curvarure is called $\rm K3$. As $\rm K3$ is a compact manifold, the spin 1/2 and spin 3/2 indexes are given by 
\bea
I_{1/2} &=& \frac{1}{768\pi^2} \int_{K3} \epsilon^{\mu\nu\rho\sigma} R_{\mu\nu\alpha\beta} R^{\alpha\beta}_{~~\rho\sigma} = -2\\
I_{3/2} &=& \frac{-21}{768\pi^2} \int_{K3} \epsilon^{\mu\nu\rho\sigma} R_{\mu\nu\alpha\beta} R^{\alpha\beta}_{~~\rho\sigma}  = + 42\\
\eea

For the ${\cal Z}_N$ symmetry, generated by $\alpha = \omega_N \equiv  \exp(i2\pi/N)$, we have 
\bea
Z \rightarrow Z e^{i (\delta S)_N}
\eea
where, 
\beq
(\delta S)_N = \frac{2\pi}{N} \left(2 \sum_{\text{spin 1/2}} q_i - 42 \sum_{\text{spin 3/2}} q_j \right) 
\label{K3-variation}
\eeq

The anomaly cancellation thus requires that the quantity within the parentheses be integer multiple of N. 

Going back to the Baryon and Lepton number symmetries, clearly the Baryon number does not have gravitational 
anomaly because the sum of baryon numbers over all the spin-1/2 Weyl fermions in the SM is zero. However, this is 
not the case for Lepton number, and it has a gravitational anomaly with $2 \sum q_i = 6$ in the SM. It then follows that $Z_2$, 
$Z_3$ and $Z_6$ subgroups of Lepton number are non-anomalous.

\subsection{Eguchi-Hanson Instanton}

Eguchi-Hanson (EH) Instanton is a solution of the vacuum (Euclidean) Einstein equation which is asymptotically 
flat and has a self dual metric. It turns out that there is no normalizable spin 1/2 zero mode in the background of 
gravitational instantons that are non-compact with $R_{\mu}^\mu \geq 0$ \cite{Gibbons:1979xn}. However, the 
existence of spin 3/2 zero modes are not excluded. 
This means that there will be no charge violation in the background of these instantons in non-supersymmetric 
theories. However, in supersymmetric theories that is not necessarily the case because of the existence of spin 3/2 
gravitino field.   

For the Eguchi-Hanson Instanton, the spin 1/2 and spin 3/2 indices were calculated in 
\cite{Eguchi:1978xp,Eguchi:1978gw,Hawking:1979zs}. They are given by,  
\begin{align}
I_{1/2} &= -\frac{1}{768\pi^2} \int_M \epsilon^{\mu\nu\rho\sigma} R_{\mu\nu\alpha\beta} R^{\alpha\beta}_{~~\rho\sigma}  +  
\frac{1}{768\pi^2} \int_{\del M} d^3x ~\epsilon^{ijk} \theta_{imn} R_{jk}^{~~mn}   - \frac{1}{2} \eta_{1/2}  - \frac{1}{2} h_{1/2} \nn \\
&= \frac{1}{8} - 0 - \frac{1}{2} \left(\frac{1}{4}\right) - 0 = 0  \label{EH-instanton-1} \\
I_{3/2} &= \frac{(21)}{768\pi^2} \int_M \epsilon^{\mu\nu\rho\sigma} R_{\mu\nu\alpha\beta} R^{\alpha\beta}_{~~\rho\sigma} - 
\frac{(21)}{768\pi^2} \int_{\del M} d^3x ~\epsilon^{ijk} \theta_{imn} R_{jk}^{~~mn}  - \frac{1}{2} \eta_{3/2} - \frac{1}{2} h_{3/2} \nn \\
&= \frac{-21}{8} - 0 - \frac{1}{2} \left(\frac{-5}{4}\right) - 0  = -2 \label{EH-instanton-2}
\end{align}

Hence, there is no spin 1/2 zero mode in this case. 
From Eq.~\eqref{EH-instanton-1} and \eqref{EH-instanton-2}  one gets 
\bea
(\delta S)_N = \frac{2\pi}{N}\sum_{ \text{spin 3/2}} (-2q)
\eea

Notice the difference with Eq.~\eqref{K3-variation} for the $K3$ instanton.
Anomaly cancellation in the background of Eguchi-Hanson Instanton thus requires that $\sum (2q)$  be integer multiple of N.

It is interesting to notice that, in ${\cal N}=1$ supergravity in 4 space-time dimensions,  the global $U(1)$ ${\cal R}$-symmetry seems to be always 
violated in the background of EH instantons irrespective of the matter content of the theory (because there are no spin-1/2 zero modes). 
We are however not able to quantify the rate of violation of the global $U(1)$ ${\cal R}$-symmetry (i.e., compute the coefficient of the corresponding 't 
Hooft operator).

\subsection{Green-Schwarz mechanism for anomaly cancellation}
The Green-Schwarz anomaly cancellation mechanism was first discovered by Green and Schwarz in the context of 
10 dimensional supergravity theories \cite{Green:1984sg}. The four-dimensional version of this mechanism works by 
invoking the existence of an axion like field (denoted by $a(x)$ below) with Wess-Zumino type interaction terms, 
\bea
{\mathcal L}_{\rm GS} \subset  \dfrac{a(x)}{f_a} \sum_i \kappa_i \, \dfrac{\alpha_i}{4\pi}\, F_{\mu\nu}^i \tilde{F}^{i\, \mu\nu}
\eea
where, the index $i$ runs over all the gauge groups and $\kappa_i$ are some numerical constants (often called the 
Kac-Moody levels that depend on how the groups $SU(3)$, $SU(2)$ and $U(1)_Y$ are embedded in a Grand Unified group).
The field $a(x)$ is assigned the following transformation property under the $U(1)$ group whose mixed anomalies are to 
be cancelled, 
\bea
\label{axion_var}
a(x) \to a(x) - \beta(x) \, f_a \, \delta_{\rm GS}
\eea
where, $\beta(x)$ is the $U(1)$ symmetry transformation parameter and $\delta_{\rm GS}$ is a constant. 
Now, note that the anomalous variation of the path integral measure under the $U(1)$ symmetry leads to the following variation 
in the Lagrangian density,
 \bea
 \label{anom_var}
\Delta {\mathcal L}  = \beta(x) \sum_i  {\mathcal A}_i \, \dfrac{\alpha_i}{4\pi} \, F_{\mu\nu}^i \tilde{F}^{i\, \mu\nu}
\eea
where, ${\mathcal A}_i$ are the anomaly coefficients. Thus, in the presence of ${\mathcal L}_{WZ}$, the total variation 
would be,
\bea
\Delta {\mathcal L}_{\rm total}  &=&  \beta(x) \left( \kappa_i \, \delta_{\rm GS} - {\mathcal A}_i \right) \, \dfrac{\alpha_i}{4\pi} \, F_{\mu\nu}^i \tilde{F}^{i\, \mu\nu} \\
&=&  \beta(x) \left(  \delta_{\rm GS} - \dfrac{{\mathcal A}_i}{\kappa_i} \right) \, \kappa_i \, \dfrac{\alpha_i}{4\pi} \,  F_{\mu\nu}^i \tilde{F}^{i\, \mu\nu}
\eea
In case of the $Z_N$ subgroup, it reduces to 
\bea
\Delta {\mathcal L}_{\rm total}  =  \dfrac{2\pi}{N} \left(  \delta_{\rm GS} - \dfrac{{\mathcal A}_i}{\kappa_i} \right) \, \kappa_i \, 
\dfrac{\alpha_i}{4\pi} \, F_{\mu\nu}^i \tilde{F}^{i\, \mu\nu}
\eea
Hence, the cancellation of all the mixed anomalies require
\bea
\delta_{\rm GS} - \dfrac{{\mathcal A}_i}{\kappa_i} = (\rm Integer) \times N , \, \, \forall \, i \, \, .
\eea
Thus, in order for this mechanism to work the quantities ${\mathcal A}_1/\kappa_1$, ${\mathcal A}_2/\kappa_2$ and ${\mathcal A}_3/\kappa_3$ 
must be equal to each other module a number which is an integer times $N$. However, note that this mechanism requires a light axion field in 
the low energy theory which may have phenomenological and cosmological implications. So we will not consider the possibility of Green-Schwarz 
anomaly cancellation in this paper.

\section{Examples}
\label{examples}

\subsection{MSSM at the renormalizable level}

In this section we present the results for various different cases. We start with the case of $Z_2$ symmetry in table~\ref{Z2-rpc}. 
The discrete $Z_2$ charges of the superspace $\theta$ coordinate and the MSSM superfields are shown in the first 8 columns. 
Note that, the discrete ${\mathcal R}$-charge of the $\theta$ coordinate (and hence also of the gauginos and gravitino), 
$Z_2(\theta,  \, \lambda, \, {\tilde G})$, is $-1$ in this case. The charge assignment $Z_2(\theta,  \, \lambda, \, {\tilde G}) = 1$ 
corresponds to non-${\mathcal R}$ discrete symmetries. 
We first look for solutions that allow all the terms in the superpotential $W_0$, but do not allow the terms in $W_{\slashed{L1}}$, 
$W_{\slashed{L2}}$, $W_{\slashed{L3}}$ and $W_{\slashed{B}}$. The discrete $Z_2$ charges of the MSSM superfields are 
obtained by imposing the requirement that the $Z_2$ symmetry does not have any quantum anomaly in the presence of 
background $SU(2)$ or $SU(3)$ gauge fields. Note that, we have not asked for the mixed anomaly cancellation with Hypercharge 
because of the ambiguities discussed in section~\ref{gauge-anomaly}. At this stage we do not check whether there exists any anomaly in the 
gravitational background. The existence of gravitational anomalies in the backgrounds of K3 and Eguchi-Hanson instantons are shown separately 
in the final two columns. 

It can be noticed that the solutions obtained for $Z_2(\theta,  \, \lambda, \, {\tilde G}) = 1$ and $-1$ are the same in table~\ref{Z2-rpc} (i.e., 
row 5 = row 7, and row 6 = row 8). 
This happens only in the $Z_2$ case, and can be easily understood by noting that 1) the charge of $\int d^2\theta$ is unity for both 
$Z_2(\theta) = 1$ and $-1$,  2) the gaugino contributions to the ${\mathcal R}$-$SU(3)$-$SU(3)$ and ${\mathcal R}$-$SU(2)$-$SU(2)$ anomalies 
are integer multiples of 2 (see Eqs.~\eqref{su3-anomaly} and \eqref{su2-anomaly}). 

It can be seen that there exists two non-anomalous sets of solutions. One of them is the standard ${\mathcal R}$-parity assignments used 
extensively in the literature. However, there also exists another set of charge assignments different from the standard ${\mathcal R}$-parity. 
An interesting question is whether these two choices are physically distinguishable. We do not have a good answer to this question. 

We have checked that there is no solution possible if any of $W_{\slashed{L1}}$ or $W_{\slashed{L2}}$ or 
$W_{\slashed{L3}}$ or  $W_{\slashed{B}}$ is included with $W_0$.

\subsubsection{$Z_2$ symmetry : }
\label{z2}

\captionsetup{width=15cm}
\renewcommand*{\arraystretch}{1.4}
\begin{center}
\begin{longtable}{c|c|c|c|c|c|c|c||c|c}
\hline \multicolumn{10}{c}{$Z_{2}$} \\ \hline
\multicolumn{10}{c}{$W_0  \,\checkmark\, , \,              	W_{\slashed{L}1}\,\times\, , \, W_{\slashed{L}2}  \,\times    \, , \, W_{\slashed{L}3} \,\times  \, , \,                      W_{\slashed{B}} \,\times $} \\ 
\hline\hline
\multicolumn{8}{c||}{Discrete charges} &                     \multicolumn{2}{c}{Free from Gravitational anomaly ?} \\          \cline{1-10}  $\theta, \, \lambda, \, {\tilde G}$ & $Q$ &         $U^c$ & $D^c$ & $L$ &  $E^c$  &  $H_u$  & $H_d$   &               K3 background & EH background \\
\hline\hline
$\hphantom{-}1$ &$\hphantom{-}1$ &$-1$ &$-1$ &$\hphantom{-}1$ &$-1$ &$-1$ &$-1$ &\checkmark &\checkmark  \\
 \hline 
$\hphantom{-}1$ &$-1$ &$-1$ &$-1$ &$-1$ &$-1$ &$\hphantom{-}1$ &$\hphantom{-}1$ &\checkmark &\checkmark  \\
 \hline 
$-1$ &$\hphantom{-}1$ &$-1$ &$-1$ &$\hphantom{-}1$ &$-1$ &$-1$ &$-1$ &\checkmark &\checkmark  \\
 \hline 
$-1$ &$-1$ &$-1$ &$-1$ &$-1$ &$-1$ &$\hphantom{-}1$ &$\hphantom{-}1$ &\checkmark &\checkmark  \\
 \hline 
\caption{The columns $1 - 8$ show the discrete $Z_2$ charges of the superspace $\theta$ coordinate as well as the MSSM superfields.  
The discrete charge $Z_2(\theta,  \, \lambda, \, {\tilde G}) = 1 \, (-1)$ of the $\theta$ coordinate, gauginos and gravitino corresponds to 
a non-${\mathcal R}$ discrete symmetry 
(discrete ${\mathcal R}$-symmetry). The charges of the MSSM superfields are obtained by demanding that there are no mixed 
anomalies of the form $Z_2 - SU(N) - SU(N)$ ($N$ = 2, 3), and subject to the constraint that only the terms in $W_0$ are allowed.  
The columns 9 and 10 show whether the discrete symmetry (defined by the charges given in columns $1 - 8$) is expected to be 
violated by gravitational anomalies in the backgrounds of K3 and Eguchi-Hanson instantons respectively. Note that, the solutions obtained 
for $Z_2(\theta,  \, \lambda, \, {\tilde G}) = 1$ and $-1$ are the same in this case.  See text for more details. \label{Z2-rpc}}
\end{longtable}
\end{center}

\vspace*{-19mm}
\subsubsection{$Z_3$ symmetry : }

In this section, we compute the various allowed $Z_3$ symmetries. Note that, for a $Z_N$ ${\mathcal R}$-symmetry 
the discrete charge for the $\theta$ coordinate can take (N-1) values : ${\rm Exp} [ i (2\pi/N)\, q_\theta ]$  ($q_\theta \in 1,...N-1$) 
which measures the difference in the discrete charges between the scalar and fermion components of a superfield. 

In Table-\ref{Z-3-1111}, we show that allowed $Z_3$ charges when only the terms in $W_0$ are allowed. It can be seen that in all the 
cases the $Z_3$ charge of the superspace $\theta$ coordinate is 1, i.e., no $Z_3$ ${\mathcal R}$ symmetry is possible. In fact, 
we do not find any non-anomalous $Z_3$ ${\mathcal R}$ symmetry at all. 

Unlike the $Z_2$ case, there are $Z_3$ symmetries which allow either only the $\slashed{B}$ term (see Table-\ref{Z-3-1112} below) 
or only the $\slashed{L}$ terms (see Table-\ref{Z-3-2221} below). This can have interesting implications for model building. For example, 
only Lepton number violating or only Baryon number violating term(s) have often been considered in the literature in a completely 
phenomenological way without any symmetry arguments (see for example, \cite{Bhattacharyya:1996nj,Biswas:2013hfa,Bardhan:2016txk,
Bardhan:2016gui,Dercks:2017lfq} and the references therein). Our results show that such choices do not always need to be ad hoc, 
and can be argued to occur based on symmetries.  

Note that, from a given set of allowed charge assignments, the replacements $\omega \to \omega^2$ and $\omega^2 \to \omega$ always 
generate another allowed solution. 
\captionsetup{width=15cm}
\renewcommand*{\arraystretch}{1.4}
\begin{center}
\begin{longtable}{c|c|c|c|c|c|c|c||c|c}
\hline \multicolumn{10}{c}{$Z_{3}$} \\ \hline
\multicolumn{10}{c}{$W_0  \,\checkmark\, , \,              	W_{\slashed{L}1}\,\times\, , \, W_{\slashed{L}2}  \,\times    \, , \, W_{\slashed{L}3} \,\times  \, , \,                      W_{\slashed{B}} \,\times $} \\ 
\hline\hline
\multicolumn{8}{c||}{Discrete charges} &                     \multicolumn{2}{c}{Free from Gravitational anomaly ?} \\          \cline{1-10}  $\theta, \, \lambda, \, {\tilde G}$ & $Q$ &         $U^c$ & $D^c$ & $L$ &  $E^c$  &  $H_u$  & $H_d$   &               K3 background & EH background \\
\hline\hline
1 &1 &$\omega$ &$\omega^2$ &1 &$\omega^2$ &$\omega^2$ &$\omega$ &\texttimes &\checkmark  \\
 \hline 
1 &1 &$\omega$ &$\omega^2$ &$\omega^2$ &1 &$\omega^2$ &$\omega$ &\texttimes &\checkmark  \\
 \hline 
1 &$\omega$ &1 &$\omega$ &1 &$\omega^2$ &$\omega^2$ &$\omega$ &\texttimes &\checkmark  \\
 \hline 
1 &$\omega$ &1 &$\omega$ &$\omega^2$ &1 &$\omega^2$ &$\omega$ &\texttimes &\checkmark  \\
 \hline 
1 &$\omega$ &$\omega$ &1 &1 &$\omega$ &$\omega$ &$\omega^2$ &\checkmark &\checkmark  \\
 \hline 
1 &$\omega$ &$\omega$ &1 &$\omega$ &1 &$\omega$ &$\omega^2$ &\checkmark &\checkmark  \\
 \hline 
\caption{See the caption of Table-\ref{Z2-rpc} for  notations, and refer to the main text for more details. 
Solutions which can be obtained by the replacements $\omega \to \omega^2$ and $\omega^2 \to \omega$ 
are removed from the table. 
\label{Z-3-1111}}
\end{longtable}
\end{center}

\captionsetup{width=15cm}
\renewcommand*{\arraystretch}{1.4}
\begin{center}
\begin{longtable}{c|c|c|c|c|c|c|c||c|c}
\hline \multicolumn{10}{c}{$Z_{3}$} \\ \hline
\multicolumn{10}{c}{$W_0  \,\checkmark\, , \,              	W_{\slashed{L}1}\,\times\, , \, W_{\slashed{L}2}  \,\times    \, , \, W_{\slashed{L}3} \,\times  \, , \,                      W_{\slashed{B}} \,\checkmark $} \\ 
\hline\hline
\multicolumn{8}{c||}{Discrete charges} &                     \multicolumn{2}{c}{Free from Gravitational anomaly ?} \\          \cline{1-10}  $\theta, \, \lambda, \, {\tilde G}$ & $Q$ &         $U^c$ & $D^c$ & $L$ &  $E^c$  &  $H_u$  & $H_d$   &               K3 background & EH background \\
\hline\hline
1 &1 &1 &1 &$\omega$ &$\omega^2$ &1 &1 &\checkmark &\checkmark  \\
 \hline 
1 &$\omega$ &$\omega^2$ &$\omega^2$ &$\omega$ &$\omega^2$ &1 &1 &\texttimes &\checkmark  \\
 \hline 
1 &$\omega$ &$\omega^2$ &$\omega^2$ &$\omega^2$ &$\omega$ &1 &1 &\texttimes &\checkmark  \\
 \hline 
\caption{See the caption of Table-\ref{Z2-rpc} for  notations, and refer to the main text for more details. 
Solutions which can be obtained by the replacements $\omega \to \omega^2$ and $\omega^2 \to \omega$ 
are removed from the table.
\label{Z-3-1112}}
\end{longtable}
\end{center}

\captionsetup{width=15cm}
\renewcommand*{\arraystretch}{1.4}
\begin{center}
\begin{longtable}{c|c|c|c|c|c|c|c||c|c}
\hline \multicolumn{10}{c}{$Z_{3}$} \\ \hline
\multicolumn{10}{c}{$W_0  \,\checkmark\, , \,              	W_{\slashed{L}1}\,\checkmark\, , \, W_{\slashed{L}2}  \,\checkmark    \, , \, W_{\slashed{L}3} \,\checkmark  \, , \,                      W_{\slashed{B}} \,\times $} \\ 
\hline\hline
\multicolumn{8}{c||}{Discrete charges} &                     \multicolumn{2}{c}{Free from Gravitational anomaly ?} \\          \cline{1-10}  $\theta, \, \lambda, \, {\tilde G}$ & $Q$ &         $U^c$ & $D^c$ & $L$ &  $E^c$  &  $H_u$  & $H_d$   &               K3 background & EH background \\
\hline\hline
1 &1 &$\omega$ &$\omega^2$ &$\omega$ &$\omega$ &$\omega^2$ &$\omega$ &\texttimes &\checkmark  \\
 \hline 
1 &$\omega$ &1 &$\omega$ &$\omega$ &$\omega$ &$\omega^2$ &$\omega$ &\texttimes &\checkmark  \\
 \hline 
1 &$\omega$ &$\omega$ &1 &$\omega^2$ &$\omega^2$ &$\omega$ &$\omega^2$ &\checkmark &\checkmark  \\
 \hline 
\caption{See the caption of Table-\ref{Z2-rpc} for  notations, and refer to the main text for more details. 
Solutions which can be obtained by the replacements $\omega \to \omega^2$ and $\omega^2 \to \omega$ 
are removed from the table.
\label{Z-3-2221}}
\end{longtable}
\end{center}

\vspace*{-15mm}
\subsubsection{$Z_4$ symmetry : }
\label{z4}

We show results for the allowed $Z_4$ symmetries in Tables - \ref{Z-4-1111}, \ref{Z-4-1112} and \ref{Z-4-2221}.
Note that the solutions for the discrete charges of the MSSM superfields for $Z_N(\theta) = {\rm Exp}[i (2\pi/N)m_1]$ and 
$Z_N(\theta) = {\rm Exp}[i (2\pi/N)m_2]$  $(m_1,m_2 \in 0,..N-1)$ are the same  if $(4\pi/N)m_1= (4\pi/N)m_2  \pm 2\pi n \, (n \in {\rm Integers})$. 
This is because, for both of these cases $d^2 \theta$ have the same transformation properties under the $Z_N$ symmetry.  Moreover, 
the anomaly constraints also do not get affected. 
For example, in the case of $Z_4$ symmetry, the solutions for $m_1 =1(Z_N(\theta) = i)$  and $m_2 =3(Z_N(\theta) = -i)$  are the same. 
Similarly,  the solutions for $m_1 =0(Z_N(\theta) = 1)$  and $m_2 =2(Z_N(\theta) = -1)$  are the same. 
Hence, we do not show the solutions for $Z_N(\theta) = -1 \text{ and } -i$.
It should also be noticed that, form a given set of allowed charge assignments, the replacement $i \to -i$ always generates another 
allowed solution. 

\captionsetup{width=  15cm}
\renewcommand*{\arraystretch}{1.4}
\begin{center}
\begin{longtable}{c|c|c|c|c|c|c|c||c|c}
\hline \multicolumn{10}{c}{$Z_{4}$} \\ \hline
\multicolumn{10}{c}{$W_0  \,\checkmark\, , \,              	W_{\slashed{L}1}\,\times\, , \, W_{\slashed{L}2}  \,\times    \, , \, W_{\slashed{L}3} \,\times  \, , \,                      W_{\slashed{B}} \,\times $} \\ 
\hline\hline
\multicolumn{8}{c||}{Discrete charges} &                     \multicolumn{2}{c}{Free from Gravitational anomaly ?} \\          \cline{1-10}  $\theta, \, \lambda, \, {\tilde G}$ & $Q$ &         $U^c$ & $D^c$ & $L$ &  $E^c$  &  $H_u$  & $H_d$   &               K3 background & EH background \\
\hline\hline
\FZ &\FTh &\FTw &\FZ &\FTh &\FZ &\FTh &\FO &\checkmark &\checkmark  \\
 \hline 
\FZ &\FZ &\FO &\FTh &\FZ &\FTh &\FTh &\FO &\texttimes &\checkmark  \\
 \hline 
\FZ &\FO &\FO &\FO &\FO &\FO &\FTw &\FTw &\texttimes &\checkmark  \\
 \hline 
\FZ &\FTw &\FTh &\FO &\FTw &\FO &\FTh &\FO &\texttimes &\checkmark  \\
 \hline 
\FZ &\FTh &\FO &\FO &\FTh &\FO &\FZ &\FZ &\texttimes &\checkmark  \\
 \hline 
\FO &\FZ &\FO &\FO &\FZ &\FO &\FO &\FO &\texttimes &\texttimes  \\
 \hline 
\FO &\FZ &\FTh &\FTh &\FZ &\FTh &\FTh &\FTh &\texttimes &\texttimes  \\
 \hline 
\FO &\FO &\FO &\FTh &\FO &\FTh &\FZ &\FTw &\texttimes &\texttimes  \\
 \hline 
\FO &\FO &\FTh &\FO &\FO &\FO &\FTw &\FZ &\texttimes &\texttimes  \\
 \hline 
\FO &\FTw &\FO &\FO &\FTw &\FO &\FTh &\FTh &\texttimes &\texttimes  \\
 \hline 
\FO &\FTw &\FTh &\FTh &\FTw &\FTh &\FO &\FO &\texttimes &\texttimes  \\
 \hline 
\FO &\FTh &\FO &\FTh &\FTh &\FTh &\FTw &\FZ &\texttimes &\texttimes  \\
 \hline 
\FO &\FTh &\FTh &\FO &\FTh &\FO &\FZ &\FTw &\texttimes &\texttimes  \\
 \hline 
 \caption{See the caption of Table-\ref{Z2-rpc} for notations, and refer to the text in section~\ref{z4} for more details. 
\label{Z-4-1111}}
\end{longtable}
\end{center}
\vspace{-1cm}
\captionsetup{width=  15cm}
\renewcommand*{\arraystretch}{1.4}
\begin{center}
\begin{longtable}{c|c|c|c|c|c|c|c||c|c}
\hline \multicolumn{10}{c}{$Z_{4}$} \\ \hline
\multicolumn{10}{c}{$W_0  \,\checkmark\, , \,              	W_{\slashed{L}1}\,\times\, , \, W_{\slashed{L}2}  \,\times    \, , \, W_{\slashed{L}3} \,\times  \, , \,                      W_{\slashed{B}} \,\checkmark $} \\ 
\hline\hline
\multicolumn{8}{c||}{Discrete charges} &                     \multicolumn{2}{c}{Free from Gravitational anomaly ?} \\          \cline{1-10}  $\theta, \, \lambda, \, {\tilde G}$ & $Q$ &         $U^c$ & $D^c$ & $L$ &  $E^c$  &  $H_u$  & $H_d$   &               K3 background & EH background \\
\hline\hline
\FO &\FZ &\FTw &\FZ &\FZ &\FZ &\FZ &\FTw &\checkmark &\texttimes  \\
 \hline 
\FO &\FO &\FTw &\FTw &\FO &\FTw &\FTh &\FTh &\checkmark &\texttimes  \\
 \hline 
\FO &\FTw &\FTw &\FZ &\FTw &\FZ &\FTw &\FZ &\checkmark &\texttimes  \\
 \hline 
\FO &\FTh &\FTw &\FTw &\FTh &\FTw &\FO &\FO &\checkmark &\texttimes  \\
 \hline 
 \caption{See the caption of Table-\ref{Z2-rpc} for  notations, and refer to the main text for more details. 
\label{Z-4-1112}}
\end{longtable}
\end{center}

\vspace{-2cm}
\captionsetup{width=  15cm}
\renewcommand*{\arraystretch}{1.4}
\begin{center}
\begin{longtable}{c|c|c|c|c|c|c|c||c|c}
\hline \multicolumn{10}{c}{$Z_{4}$} \\ \hline
\multicolumn{10}{c}{$W_0  \,\checkmark\, , \,              	W_{\slashed{L}1}\,\checkmark\, , \, W_{\slashed{L}2}  \,\checkmark    \, , \, W_{\slashed{L}3} \,\checkmark  \, , \,                      W_{\slashed{B}} \,\times $} \\ 
\hline\hline
\multicolumn{8}{c||}{Discrete charges} &                     \multicolumn{2}{c}{Free from Gravitational anomaly ?} \\          \cline{1-10}  $\theta, \, \lambda, \, {\tilde G}$ & $Q$ &         $U^c$ & $D^c$ & $L$ &  $E^c$  &  $H_u$  & $H_d$   &               K3 background & EH background \\
\hline\hline
\FO &\FZ &\FZ &\FTw &\FZ &\FTw &\FTw &\FZ &\checkmark &\texttimes  \\
 \hline 
\FO &\FO &\FZ &\FZ &\FO &\FZ &\FO &\FO &\checkmark &\texttimes  \\
 \hline 
\FO &\FTw &\FZ &\FTw &\FTw &\FTw &\FZ &\FTw &\checkmark &\texttimes  \\
 \hline 
\FO &\FTh &\FZ &\FZ &\FTh &\FZ &\FTh &\FTh &\checkmark &\texttimes  \\
 \hline 
\caption{See the caption of Table-\ref{Z2-rpc} for  notations, and refer to the main text for more details. 
\label{Z-4-2221}}
\end{longtable}
\end{center}


\vspace*{-19mm}
\subsection{MSSM including higher dimensional operators}

In table~\ref{hdo}, we show the list of gauge-invariant super-potential operators in the MSSM of mass dimensions 2, 3, 4, 5 and 6 in superspace. 
We do not show the gauge and family indices of the operators. 

\begin{table}
\begin{center}
{\scriptsize
\begin{eqnarray*}
\hspace*{1.5cm}\begin{array}{|c|c|c|c|c|}
\hline
\multicolumn{5}{|c|}{D = 2}\\
\hline
 & \text{Operator} & \text{B} & \text{L} & \text{Comment}\\
\hline
 {\cal O}^2_1 & H_u H_d & 0 & 0 & \\
 {\cal O}^2_2 & L H_u & 0 & 1 & \\
 \hline
\multicolumn{5}{|c|}{D = 3}\\
\hline
 {\cal O}^3_1 & Q U^c H_u & 0 & 0 & \\
 {\cal O}^3_2 &Q D^c H_d & 0 & 0 & \\
 {\cal O}^3_3 &L E^c H_d & 0 & 0 & \\
 {\cal O}^3_4 &L L E^c & 0 & 1 & \\
 {\cal O}^3_5 &Q D^c L & 0 & 1 &\\
 {\cal O}^3_6 &U^c D^c D^c & -1 & 0 & \\
 \hline
\multicolumn{5}{|c|}{D = 4}\\
\hline
 {\cal O}^4_1 &Q Q U^c D^c & 0 & 0  & \\
 {\cal O}^4_2 &Q U^c L E^c & 0 & 0  & \\
 {\cal O}^4_3 &H_u H_u H_d H_d & 0 & 0 & \left({\cal O}^2_1\right)^2\\
 {\cal O}^4_4 &Q U^c E^c H_d & 0 & -1  & \\
 {\cal O}^4_5 &L L H_u H_u & 0 & 2  & \left({\cal O}^2_2\right)^2\\
 {\cal O}^4_6 &L H_u H_u H_d & 0 & 1  & {\cal O}^2_1 {\cal O}^2_2\\ 
 {\cal O}^4_7 &Q Q Q H_d & 1 & 0  & \\
 {\cal O}^4_8 & Q Q Q L & 1 & 1  & \\
 {\cal O}^4_9 &U^c U^c D^c E^c & -1 & -1  & \\
 \hline
\multicolumn{5}{|c|}{D = 5}\\
\hline 
 {\cal O}^5_1 & Q U^c H_u H_u H_d & 0 & 0 & {\cal O}^3_1 {\cal O}^2_1 \\
 {\cal O}^5_2 & Q D^c H_u H_d H_d & 0 & 0 & {\cal O}^3_2 {\cal O}^2_1 \\
 {\cal O}^5_{3} & L E^c H_u H_d H_d & 0 & 0 & {\cal O}^3_3 {\cal O}^2_1\\ 
 {\cal O}^5_4 & Q Q U^c U^c E^c & 0 & -1 & \\
 {\cal O}^5_5 & Q U^c L H_u H_u & 0 & 1 & {\cal O}^3_1 {\cal O}^2_2\\ 
 {\cal O}^5_6 & Q D^c L L H_u & 0 & 2 & {\cal O}^3_5 {\cal O}^2_2\\ 
 {\cal O}^5_7 & Q D^c L H_u H_d & 0 & 1 & {\cal O}^3_5 {\cal O}^2_1\\ 
 {\cal O}^5_{8} & L L L E^c H_u & 0 & 2 & {\cal O}^3_4 {\cal O}^2_2\\
 {\cal O}^5_{9} & L L E^c H_u H_d & 0 & 1 & {\cal O}^3_4 {\cal O}^2_1\\ 
 {\cal O}^5_{10} & E^c H_u H_d H_d H_d & 0 & -1 & \\  
 {\cal O}^5_{11} &  Q Q Q Q U^c & 1 & 0 & \\  
 {\cal O}^5_{12} & U^c D^c D^c H_u H_d & -1 & 0 & {\cal O}^3_6 {\cal O}^2_1 \\ 
 {\cal O}^5_{13} &  U^c U^c U^c E^c E^c & -1 & -2 & \\  
 {\cal O}^5_{14} &  U^c D^c D^c L H_u & -1 & 1 & {\cal O}^3_6 {\cal O}^2_2\\ 
  {\cal O}^5_{15} & D^c D^c D^c L L & -1 & 2 & \\  
 {\cal O}^5_{16} & D^c D^c D^c L H_d & -1 & 1 & \\  
 \hline
\end{array}
\hspace{8mm}
\begin{array}{|c|c|c|c|c|}
\hline
\multicolumn{5}{|c|}{D = 6}\\
\hline
& \text{Operator} & \text{B} & \text{L} & \text{Comment}\\
\hline\hline
 {\cal O}^6_1 & Q Q U^c U^c H_u H_u & 0 & 0 & \left( {\cal O}^3_1\right)^2 \\
  {\cal O}^6_2 & Q Q U^c D^c H_u H_d & 0 & 0 & {\cal O}^3_1 {\cal O}^3_2 \\
  {\cal O}^6_3 & Q Q D^c D^c H_d H_d & 0 & 0 & \left({\cal O}^3_2 \right)^2\\
  {\cal O}^6_{4} & Q U^c L E^c H_u H_d & 0 & 0 & {\cal O}^3_1 {\cal O}^3_3\\
  {\cal O}^6_{5} & Q D^c L E^c H_d H_d & 0 & 0 & {\cal O}^3_2 {\cal O}^3_3 \\
 {\cal O}^6_{6} & H_u H_u H_u H_d H_d H_d & 0 & 0 & \left({\cal O}^2_1 
\right)^2 \\    
   {\cal O}^6_7 & Q Q U^c D^c L H_u & 0 & 1 & {\cal O}^3_1 {\cal O}^3_5\\
 {\cal O}^6_8 & Q Q D^c D^c L L & 0 & 2 & \left( {\cal O}^3_5\right)^2 \\
 {\cal O}^6_9 & Q Q D^c D^c L H_d & 0 & 1 & {\cal O}^3_2 {\cal O}^3_5\\
 {\cal O}^6_{10} & Q U^c L L E^c H_u & 0 & 1 & {\cal O}^3_1 {\cal O}^3_4 \\  
 {\cal O}^6_{11} & Q U^c E^c H_u H_d H_d & 0 & -1 & \\  
 {\cal O}^6_{12} & Q D^c L L L E^c & 0 & 2 & {\cal O}^3_4{\cal O}^3_5 \\ 
 {\cal O}^6_{13} & Q D^c L L E^c H_d & 0 & 1 & {\cal O}^3_3 {\cal O}^3_5\\ 
 {\cal O}^6_{14} & Q D^c E^c H_d H_d H_d & 0 & -1 & \\ 
 {\cal O}^6_{15} & L L L L E^c E^c & 0 & 2 & \left({\cal O}^3_4\right)^2\\  
 {\cal O}^6_{16} & L L L E^c E^c H_d & 0 & 1 & {\cal O}^3_3 {\cal O}^3_4 \\
 {\cal O}^6_{17} & L L L H_u H_u H_u & 0 & 3 & \left({\cal O}^2_2\right)^3 \\  
 {\cal O}^6_{18} & L L H_u H_u H_u H_d & 0 & 2 & {\cal O}^2_1 \left( {\cal 
O}^2_2 \right)^2 \\ 
 {\cal O}^6_{19} & L E^c E^c H_d H_d H_d & 0 & -1 & \\
 {\cal O}^6_{20} & L H_u H_u H_u H_d H_d & 0 & 1 & \left({\cal O}^2_1\right)^2 
{\cal O}^2_2 \\
 {\cal O}^6_{21} & E^c E^c H_d H_d H_d H_d & 0 & -2 & \\
  {\cal O}^6_{22} & Q Q Q H_u H_d H_d & 1 & 0 & {\cal O}^4_7 {\cal O}^2_1 \\
  {\cal O}^6_{23} &  Q U^c U^c D^c D^c H_u & -1 & 0 & {\cal O}^3_1 {\cal 
O}^3_6\\ 
 {\cal O}^6_{24} & U^c U^c D^c D^c D^c D^c & -2 & 0 & \left({\cal 
O}^3_6\right)^2\\
 {\cal O}^6_{25} & U^c U^c D^c L E^c H_u & -1 & 0 & \\ 
 {\cal O}^6_{26} & U^c D^c D^c L E^c H_d & -1 & 0 & {\cal O}^3_3 {\cal 
O}^3_6\\ 
 {\cal O}^6_{27} & Q U^c D^c D^c D^c H_d & -1 & 0 & {\cal O}^3_2 {\cal O}^3_6 
\\ 
 {\cal O}^6_{28} &Q Q Q L L H_u & 1 & 2 & {\cal O}^4_8 {\cal O}^2_2\\
 {\cal O}^6_{29} & Q Q Q L H_u H_d & 1 & 1 & {\cal O}^4_1 {\cal O}^2_1 \\
  {\cal O}^6_{30} & Q U^c D^c D^c D^c L & -1 & 1 & {\cal O}^3_5 {\cal O}^3_6\\ 
 {\cal O}^6_{31} & U^c U^c D^c E^c H_u H_d & -1 & -1 & \\ 
 {\cal O}^6_{32} & U^c D^c D^c L L E^c & -1 & 1 &{\cal O}^3_4 {\cal O}^3_6 \\
 \hline
\multicolumn{5}{c}{\phantom{aa}} \\
\multicolumn{5}{c}{\phantom{aa}} \\
\multicolumn{5}{c}{\phantom{aa}} \\
\multicolumn{5}{c}{\phantom{aa}}
\end{array}
\end{eqnarray*}}
\caption{\small Gauge invariant superpotential operators of mass dimensions 2, 3, 4, 5 and 6 in the MSSM. The Baryon and Lepton numbers 
of the operators are also shown. In the ``Comment" column we show whether an operator structure is made out of product of two 
lower dimensional operators.\label{hdo}}
\end{center}
\end{table}

Note that, a superspace operator of dimension $D$ can generate operators in ordinary space of dimension up to 2(D-1).
We now check whether it is possible to remove all the Baryon number violating superspace operators up to dimension 6 by suitable $Z_N$ charge assignments. In tables.~\ref{HDO-Z3}  we show such charge assignments for $Z_3$ symmetry. 
We do not find any solution for the $Z_2$ and $Z_4$ cases.  

It can be seen that there indeed exists one completely non-anomalous $Z_3$ solution which forbids all Baryon number violation operators 
up to dimension 6 level. All the $Z_4$ solutions have gravitational anomalies. Note that, in this analysis we have ignored possible 
higher dimensional operators in the K{\"a}hler potential for simplicity.

\captionsetup{width=  15cm}
\renewcommand*{\arraystretch}{1.4}
\begin{center}
\begin{longtable}{c|c|c|c|c|c|c|c||c|c}
\hline \multicolumn{10}{c}{$Z_{3}$} \\ \hline
\hline\hline
\multicolumn{8}{c||}{Discrete charges} &                     \multicolumn{2}{c}{Free from Gravitational anomaly ?} \\          \cline{1-10}  $\theta, \, \lambda, \, {\tilde G}$ & $Q$ &         $U^c$ & $D^c$ & $L$ &  $E^c$  &  $H_u$  & $H_d$   &               K3 background & EH background \\
\hline\hline
\ThZ &\ThZ &\ThO &\ThTw &\ThO &\ThO &\ThTw &\ThO &\texttimes &\checkmark  \\
 \hline 
\ThZ &\ThO &\ThZ &\ThO &\ThO &\ThO &\ThTw &\ThO &\texttimes &\checkmark  \\
 \hline 
\ThZ &\ThO &\ThO &\ThZ &\ThTw &\ThTw &\ThO &\ThTw &\checkmark &\checkmark  \\
 \hline 
\caption{$Z_3$ symmetry solutions for the case when no $\slashed{B}$ operator is allowed up to dimension 6 level in superspace. 
Solutions which can be obtained by the replacements $\omega \to \omega^2$ and $\omega^2 \to \omega$ 
are removed from the table.
\label{HDO-Z3}}
\end{longtable}
\end{center}

\vspace*{-20mm}
\section{Summary}
\label{summary}

It is widely believed that there cannot be any global symmetry in gravity. This means that all global symmetries in a QFT are violated when 
it is coupled to gravity. However, in the context of particle physics model building, both continuous global symmetries and discrete symmetries 
have been extensively used. In particular, discrete symmetries have found a lot a applications because, even if the discrete symmetry is 
broken, no unwanted goldstone boson arises.  A well known example is ${\cal R}$-parity in the MSSM which is a $Z_2$ symmetry 
imposed ad-hoc in the Lagrangian in order to avoid fast proton decay. 

There are two ways to justify imposition of such $Z_2$, or in general $Z_N$ symmetry on the Lagrangian. The first is to assume that 
the $Z_N$ is an accidental global symmetry of the low energy Lagrangian, and does not appear in the UV theory. However, in this case, 
it is only expected to be an approximate symmetry. The second possibility is to demand that the $Z_N$ symmetry is a remnant of a 
spontaneously broken gauge symmetry in the UV. We call it a gauged discrete symmetry. 

However, in order to promote a discrete global symmetry to a gauge symmetry, it must be non-anomalous. In general, these anomaly constraints 
can be linear, quadratic or cubic in the $Z_N$ charges. Unfortunately, only those constraints which is linear in the $Z_N$ charges 
are useful, and the ones non-linear in the charges are UV dependent. In this paper, we systematically study these linear constraints taking 
into account both the gauge and known gravitational instantons. 

We apply these constraints  to the MSSM superpotential to show that \\[2mm]
-- a large class of $Z_N$ symmetries are ruled out because of the existence of mixed anomalies with the gauge symmetries. 
Among those which are free from gauge anomalies, a significant number of them is violated in the background of $\rm K3$ and Eguchi-Hanson 
gravitational backgrounds.  \\[2mm]
-- ${\cal R}$-parity indeed does not have any gauge or gravitational anomalies, hence can be considered as a gauged $Z_2$ 
symmetry (see table~\ref{Z2-rpc}). \\[2mm]
-- There exists non-anomalous discrete $Z_3$ symmetries that allow to consider only Baryon number violating or only Lepton number 
violating terms in the MSSM superpotential (see tables~\ref{Z-3-1112}, \ref{Z-3-2221}).

We also list all the gauge invariant higher dimensional superpotential operators of dimension 4, 5 and 6 in the MSSM. We show that  
there exists non-anomalous $Z_3$ symmetry (table~\ref{HDO-Z3}) which forbids all baryon number violating terms up to dimension 
6 in the superpotential. 

As a by-product of our analysis, we observe that in a theory of $N=1$ supergravity in 4 space-time dimensions, the global 
$U(1)$ ${\cal R}$-symmetry is broken by the Eguchi-Hanson gravitational instanton irrespective of the matter content of the theory. 

Before we close, we would like to briefly mention a few questions which have not been answered in this work, and deserves further investigations. 
We list them below:\\[2mm]
-- As pointed out by Witten \cite{Witten:1985xe}, it is not completely clear whether all gravitational instantons need to be included in the 
path integral, or in other words, whether excluding them from the path integral will lead to some mathematical inconsistencies. It was argued that the 
argument based on cluster decomposition \cite{Callan:1976je} and the existence of anti-instanton (for every instanton configuration), which works for gauge theories, 
does not necessarily apply in the gravitational case.  In this work, we have ignored this caveat, and assumed that, like gauge theories, instantons 
have to be included in the path integral for consistency. \\[2mm]
-- We have not explored the phenomenological implications of the various non-anomalous $Z_N$ symmetries that we have found in our analysis. 
It would interesting to investigate whether they lead to distinct experimental signatures. \\ [2mm]
-- Both K3 and Eguchi-Hanson instantons are classical solutions of the Euclidean Einstein equation with vanishing cosmological constant. 
It is not clear whether considering such solutions are consistent with the fact that the observed cosmological constant in the context of 
the Friedmann-Robertson-Walker cosmology is non-zero and positive. \\[2mm]
-- While we have talked about the violation of global symmetries by gauge and specific gravitational instantons, no quantitative understanding of the 
rate of such violation is provided. This would require computation of the 't Hooft operator and its coefficient corresponding 
to the instanton solutions. To our knowledge, even for the gauge instantons, it is not clear how to compute the coefficient of the 't Hooft operator 
in general. Hence, only the naive estimate that the rate of violation should be proportional to $e^{-S_E}$ is usually provided. 
The situation for the gravitational instantons is even less understood. Firstly, there are only a few gravitational instanton solutions are known. 
Moreover, no calculation of the 't Hooft operator in the gravitational case exists.

\subsection*{Acknowledgement}
We thank Ofer Aharony, Zakaria Chacko, Hooman Davoudiasl, Zohar Komargodski, Rabindra Nath Mohapatra, Shiraz Minwalla,  Lorenzo Di Pietro, 
Nathan Seiberg, Ashoke Sen and Raman Sundrum for discussions.

\vspace*{5mm}
\appendix
\begin{center}
\Large{{\bf Appendix}}
\end{center}

\section{Anomalies and Index \label{Fujikawa}} 

In this section we review, following Fujikawa \cite{Fujikawa:1983bg}, the anomalous variation 
of the fermion measure under a chiral $U(1)$ 
transformation, to show that it is equal to the Index of the covariant Dirac operator. 
We will concentrate only on the fermion path integral and treat the coupled gauge 
(or gravitational) field as a fixed background field. The relevant path-integral is 
\beq\begin{split}\label{pidef}
Z &= \int D\Psi D\bar\Psi \exp\left( \int d^4x \, \bar\Psi \gamma^\mu iD_\mu \Psi \right)
\end{split}\eeq 
where $iD_\mu= i\del_\mu + A_\mu + \omega_\mu$, $\omega_\mu$ being the spin 
connection.

The chiral $U(1)$ symmetry acts on the Dirac fermion $\Psi$ as 
\beq
\begin{split}
\Psi(x) \longrightarrow \Psi'(x) &= e^{i q \alpha(x) \gamma_5} \Psi(x) ~~~ 
       \left( \text{and consequently}, \bar\Psi(x) \longrightarrow \bar\Psi'(x) 
    = \bar\Psi(x) e^{i q \alpha(x) \gamma_5} \right)\\ 
& = \int d^4y \, \delta^4(y-x) \, e^{i q \alpha(y) \gamma_5} \Psi(y) \\
& = \int d^4y \, J(x,y) \, \Psi(y) \\
\textrm{where} & \quad J(x,y)= e^{i q \alpha(y) \gamma_5}\delta^4(y-x)~
\textrm{is the matrix of transformations.}\\ 
\textrm{Similarly, } \quad \bar{\Psi}'(x) &= \int d^4y \, \bar\Psi(y) \, J(x,y) \, .
\end{split} 
\eeq 

Notice that, in the transformation above the Dirac space transformation matrix is the same for both 
$\Psi$ and $\bar\Psi$. Under these transformations the measure then changes
as follows (taking Grassman nature of fermion into account) 
\beq\begin{split}
D\Psi D\bar\Psi & = (\det J)^2 \, D\Psi' D\bar\Psi' \\
 &= e^{2\log \det(J)} \, D\Psi' D\bar\Psi' \\
 &= e^{2 \rm{Tr} (\log J)} \, D\Psi' D\bar\Psi' \\
\end{split}\eeq
The `$\det$' and `tr' above are over the Dirac space as well as the space time, and are formally 
defined as follow
\beq\begin{split}\label{trlogJ}
2 \rm{Tr} \log J &= \rm{tr}_D \int d^4x ~d^4y ~ \delta^4(x-y) ~ [(\log J)(x,y)] \\ 
\rm{using } \quad & [\log J](x,y)= \log(e^{i q \alpha\gamma_5}) ~ \delta^4(x-y)  
  = i q \alpha \gamma_5  ~ \delta^4(x-y), \quad \textrm{we have} \\
2 \rm{Tr} \log J &= \rm{Tr}_D \int d^4x~d^4y ~ \delta^4(x-y) ~ \delta^4(x-y) (i\alpha\gamma_5) \\
  &= 2iq ~\rm{Tr}_D \int d^4x \left[ \alpha\gamma_5 ~ \delta^4(x-x) \right] \\
\end{split}\eeq 
In the above expressions `$\rm{Tr}$' is a trace over spacetime as well as Dirac space while 
`$\rm{Tr}_D$' is a trace only over the Dirac space.

The formal expression in the integrand is a product of a divergent quantity, $\delta^4(x-x)$, 
and a vanishing quantity, $\rm{Tr}_D(\gamma_5)$, and thus need to be regularized to make sense of. 
Since we are eventually interested in cases where the background gauge (gravitational) fields are 
dynamical, our regularization procedure must respect the corresponding gauge (diffeomorphism)
symmetries. 

A convenient basis for this purpose are the eigenvectors of the gauge (diffeomorphism) 
covariant Dirac operator, $\gamma^\mu iD_\mu$, in equation \eqref{pidef}
\beq 
i\gamma^\mu D_\mu \Psi_n(x)= \lambda_n \Psi_n(x). 
\eeq
Using the completeness relation 
\beq
I_4 \delta^4(x-y)=\sum_n \Psi_n(x) \bar{\Psi}_n(y),
\eeq
we can rewrite the Jacobian as follows 
\beq\begin{split}
{\cal A} &= 2 \rm{Tr}\log J 
       = 2iq ~\rm{Tr}_D \int d^4x ~\alpha 
          \left( \sum_n \Psi_n(x) \bar{\Psi}_n(x) \gamma_5 \right) \\
      & = 2iq ~\int d^4x ~\alpha 
          \left( \sum_n \bar{\Psi}_n(x)\gamma_5 \Psi_n(x)  \right).
\end{split}\eeq
This expression is a simple rewriting of the \eqref{trlogJ} and is also divergent. 
The usefulness of these basis though is that it suggests a natural gauge (diffeomorphism) 
preserving regularization by using the eigenvalues $\{\lambda_n\}$ and a regulator mass parameter 
$M$ as follows 
\beq\begin{split}
{\cal A}
 &= 2iq ~\int d^4x ~\alpha 
    \left( \sum_n \bar{\Psi}_n(x) e^{-\lambda_n^2/M^{2}} \gamma_5 \Psi_n(x) \right) \\
\end{split}\eeq

Since the covariant Dirac operators anti-commutes with $\gamma_5$, the eigenfunctions 
for non-vanishing eigenvalues always appear in pairs $\{ \Psi_n, \gamma_5 \Psi_n \}$ with 
opposite eigenvalues $\{ \lambda_n, -\lambda_n \}$ respectively\footnote{The same is not 
necessarily true for zero modes.}. Using this, one can rearrange the above sum to rewrite
\beq\begin{split}
{\cal A}
 &= 2iq\alpha \int d^4x \bigg[
    \sum_{\lambda_n=0} \left( \bar{\Psi}_n(x) \gamma_5 \Psi_n(x) \right) + \\
   & \qquad  \sum_{\lambda_n >0} \left( \bar{\Psi}_n(x) e^{-\lambda_n^2/M^{2}} \gamma_5 \Psi_n(x) 
    + \big(-\bar{\Psi}_n(x)\gamma_5\big) e^{-(-\lambda_n)^2/M^{2}} \gamma_5 
      \big(\gamma_5\Psi_n(x)\big) \right) \bigg], \\
\end{split}\eeq
which shows that all the non-zero modes cancel out in pair and we are only left with 
the contribution of zero modes which will in general be non-vanishing. 
\beq\begin{split}
{\cal A}
 &= 2iq\alpha \int d^4x \sum_{\lambda_n=0} \left( \bar{\Psi}_n(x) \gamma_5 \Psi_n(x) \right) \\
 &=  2iq\alpha ~(n_+ - n_-) = 2iq\alpha ~I_{1/2}. \\
\end{split}\eeq
This quantity, which the difference of normalizable left handed and right handed 
zero mdoes, is referred to as the ``{\it Index of the covariant Dirac operator}". 
The above expression generalizes in a straightforward way to the case of 
multiple Dirac fermions, $\Psi_j$, transforming in representations 
${\cal R}_j$ under the gauge group and carrying charge $q_j$ under the chiral 
transformations, one gets 
\beq
{\cal A}_D({q_j,{\cal R}_j}) =i (2\alpha) \sum_{j} q_j I_{1/2}({\cal R}_j) 
        = i (2\alpha) \sum_{j} q_j (n_{j+} - n_{j-})
\eeq
where $n_{j\pm}$ are the number of left(+)/right(-) handed zero modes of the covariant 
Dirac operators in representation ${\cal R}_j$. For the case of a general non-abelian 
gauge field\footnote{For compact gauge groups.} background the Atiyah-Singer Index theorem 
relates the above difference of zero modes to the Pontryagin number ($\nu$) for the 
background gauge field configuration 
\beq\begin{split} 
n_{j+} - n_{j-} &= - 2T({\cal R}_j) \nu \\
                     &=  -\frac{2 T({\cal R}_j)}{64\pi^2} \int d^4x ~ \epsilon^{\mu\nu\lambda\rho} 
                         F^a_{\mu\nu} F^a_{\lambda\rho} \\ 
     \rm{where}  \quad & T({\cal R}_j)\delta_{ab} = \rm{Tr}_{{\cal R}_j}(T_aT_b)
\end{split}\eeq
The Pontryagin number, $\nu$, is always an integer and for the minimal gauge 
instanton take value $\nu= 1$. 

For the case of single left handed Weyl fermion the above discussed method 
doesn't directly apply but a modification of this method 
\cite{AlvarezGaume:1983cs,AlvarezGaume:1984dr} 
give the mixed global-gauge-gauge anomaly to be just the half of the anomaly for a 
corresponding Dirac (i.e. transforming in the same representation of gauge group and 
and carrying the same global charge) fermion. 
\beq\begin{split}
{\cal A}_W(q_j,{\cal R}_j) 
	  &= \half \sum_j {\cal A}_D(q_j,{\cal R}_j) \\
          &= i\alpha \sum_j q_j I_{1/2}({\cal R}_j) \\
\end{split}\eeq

%
\providecommand{\href}[2]{#2}\begingroup\raggedright\endgroup

%
\end{document}